\documentclass[12pt]{article}
\usepackage{amsmath}

\begin{document}

\begin{titlepage}
\title{\bf Bi-Para-Mechanical Systems on Lagrangian Distributions}
\author{ Mehmet Tekkoyun \footnote{Corresponding author. E-mail address: tekkoyun@pau.edu.tr; Tel: +902582953616; Fax: +902582953593}\\
{\small Department of Mathematics, Pamukkale University,}\\
{\small 20070 Denizli, Turkey}\\
Murat Sari\\
{\small Department of Mathematics, Pamukkale University,}\\
{\small 20070 Denizli, Turkey}}
\date{\today}
\maketitle

\begin{abstract}

In this work, bi-para-complex analogue of Lagrangian and Hamiltonian
systems was introduced on Lagrangian distributions. Yet, the geometric and physical
results related to bi-para-dynamical systems were also presented.



\end{abstract}
\end{titlepage}

\section{Introduction}

In modern differential geometry, a convenient vector field defined on the
tangent and cotangent spaces (or manifolds) which are phase-spaces of
velocities and momentum of a given configuration space explains the dynamics
of Lagrangians and Hamiltonians. If $Q$ is an $m$-dimensional configuration
manifold and $L(H):TQ(T^{*}Q)\rightarrow \mathbf{R}$\textbf{\ }is a regular
Lagrangian (Hamiltonian) function, then there is a unique vector field $\xi $
($Z_{H}$) on $TQ$ ($T^{*}Q$) such that dynamical equations are given by
\begin{equation}
i_{\xi }\Phi _{L}=dE_{L},\,\,\,\,i_{Z_{H}}\Phi _{H}=dH  \label{1.1}
\end{equation}
where $\Phi _{L}(\Phi _{H})$ indicates the symplectic form. The triples $%
(TQ,\Phi _{L},\xi )$ and $(T^{*}Q,\Phi _{H},Z_{H})$ are called \textit{%
Lagrangian system} and \textit{Hamiltonian system }on the tangent bundle $TQ$
and on the cotangent bundle $T^{*}Q,$ respectively$.$

Mathematical models of mechanical systems are generally described by the
Hamiltonian and Lagrangian systems. These models, in particular geometric
models in mechanics, are given in some studies \cite{mcrampin, nutku,deleon}%
. Klein \cite{klein} showed that the geometric study of the Lagrangian
theory admits an alternative approach without the use of the regular
condition on $L$. It is possible to describe such formalism by using only
the differential geometry of the tangent bundles. This is an independent
formulation of the Hamiltonian theory. In fact, the role of tangent geometry
in Lagrangian theories can be seen to be similar to that of symplectic
geometry in Hamiltonian theories \cite{deleon}. Para-complex analogues of
the Euler-Lagrange and Hamiltonian equations were obtained in the framework
of K\"{a}hlerian manifold and the geometric results on a para-complex
mechanical systems were found \cite{tekkoyun2005}. Symplectic geometry has
important roles in some fields of mathematics and physics \cite{etayo}. As
known, Lagrangian foliations on symplectic manifolds are used in geometric
quantization and a connection on a symplectic manifold is an important
structure to obtain a deformation quantization. A para-K\"{a}hlerian
manifold $M$ is said to be endowed with an almost bi-para-Lagrangian
structure (a bi-para-Lagrangian manifold) if $M$ has two transversal
Lagrangian distributions (involutive transversal Lagrangian distributions) $%
D_{1}$ and $D_{2}$ \cite{etayo}.

In the above studies; although para-complex geometry and para-complex
mechanical systems were analyzed successfully, they have not dealt with
bi-para-complex mechanical systems on Lagrangian distributions. In this
Letter, therefore, equations related to bi-para-mechanical systems on
Lagrangian distributions used in obtaining geometric quantization have been
presented.

\section{Preliminaries}

Throughout this Letter, all mathematical objects and mappings are assumed to
be smooth, i.e. infinitely differentiable and Einstein convention of
summarizing is adopted, and also $1\leq i\leq m$. $A$, $\mathcal{F}(M)$, $%
\chi (M)$ and $\Lambda ^{1}(M)$ denote the set of para-complex numbers, the
set of para-complex functions on $M$, the set of para-complex vector fields
on $M$ and the set of para-complex 1-forms on $M$, respectively. Since a
para-manifold has two transversal Lagrangian distributions; instead of
\textit{para} term, \textit{bi-para} term is used for the structures or the
mentioned manifold. All structures on the differential manifold $M$
introduced in \cite{etayo,cruceanu} can be extended in the following
subsections.

\subsection{Bi-para-complex manifolds}

\textbf{Definition\thinspace \thinspace 1:} A \textit{product structure} $J$
on a manifold $M$ is a (1,1) tensor field $J$ on $M$ such that $J^{2}=I.$
The pair $(M,J)$ is called a \textit{product manifold}. A \textit{%
bi-para-complex manifold} $(M,J,D_{1},D_{2})$ is a product manifold $(M,J)$
and then one can define a (1,1) tensor field $J$ by $J|_{D_{1}}=I$ and $%
J|_{D_{2}}=-I$. Obviously, $J^{2}=I$ and the Nijenhuis (1,2) tensor field $%
N_{J}$ vanishes iff $D_{1}$ and $D_{2}$ distributions are involutive.

Let $(\ x^{i},\,\ y^{i})$ be a real coordinate system on a neighborhood $U$
of any point $p$ of $M,$ and let $\{(\frac{\partial }{\partial x^{i}})_{p},(%
\frac{\partial }{\partial y^{i}})_{p}\}$ and $\{(dx^{i})_{p},(dy^{i})_{p}\}$
be natural bases over $R$ of the tangent space $T_{p}(M)$ and the cotangent
space $T_{p}^{*}(M)$ of $M,$ respectively$.$ Then the definitions can be
given by
\begin{equation}
J(\frac{\partial }{\partial x^{i}})=\frac{\partial }{\partial y^{i}}%
,\,\,\,\,\,\,\,J(\frac{\partial }{\partial y^{i}})=\frac{\partial }{\partial
x^{i}}.  \label{2.1}
\end{equation}
Let $\ z^{i}=\ x^{i}+$\textbf{j}$\ y^{i},$ \textbf{j}$^{2}=1,$ also be a
para-complex local coordinate system on a neighborhood $U$ of any point $p$
of $M.$ The vector fields can then be shown:
\begin{equation}
(\frac{\partial }{\partial z^{i}})_{p}=\frac{1}{2}\{(\frac{\partial }{%
\partial x^{i}})_{p}-\mathbf{j}(\frac{\partial }{\partial y^{i}}%
)_{p}\},\,\,\,(\frac{\partial }{\partial \overline{z}^{i}})_{p}=\frac{1}{2}%
\{(\frac{\partial }{\partial x^{i}})_{p}+\mathbf{j}(\frac{\partial }{%
\partial y^{i}})_{p}\}.  \label{2.2}
\end{equation}
And the dual covector fields are:
\begin{equation}
\left( dz^{i}\right) _{p}=\left( dx^{i}\right) _{p}+\mathbf{j}%
(dy^{i})_{p},\,\,\,\,\,\left( d\overline{z}^{i}\right) _{p}=\left(
dx^{i}\right) _{p}-\mathbf{j}(dy^{i})_{p}  \label{2.3}
\end{equation}
which represent the bases of the tangent space $T_{p}(M)$ and cotangent
space $T_{p}^{*}(M)$ of $M$, respectively. Then the following expression can
be found
\begin{equation}
J(\frac{\partial }{\partial z^{i}})=\mathbf{-j}\frac{\partial }{\partial
z^{i}},\,\,\,\,\,\,J(\frac{\partial }{\partial \overline{z}^{i}})=\mathbf{j}%
\frac{\partial }{\partial \overline{z}^{i}}.  \label{2.4}
\end{equation}
The dual endomorphism $J^{*}$ of the cotangent space $T_{p}^{*}(M)$ at any
point $p$ of manifold $M$ satisfies that $J^{*2}=I,$ and is defined by
\begin{equation}
J^{*}(dz^{i})=\mathbf{-j}dz^{i},\,\,\,\,\,\,J^{*}(d\overline{z}^{i})=\mathbf{%
j}d\overline{z}^{i}.  \label{2.5}
\end{equation}

\subsection{Bi-para-complex modules and bi-para-Hermitian and bi-para-K\"{a}%
hlerian manifolds}

Here, some definitions and structures on para-Hermitian and para-K\"{a}%
hlerian\textit{\ }manifold $M$ introduced in \cite{etayo,cruceanu} are
extended to the distributions $D_{1}$ and $D_{2}.$

Let $V^{A}$ be a commutative group $(V,+)$ endowed with a structure of
unitary module over the ring $A$ of para-complex numbers. Let $V^{R}$ denote
the group $(V,+)$ endowed with the structure of real vector space inherited
from the restriction of scalars to $R\mathbf{.}$ The vector space $V^{R}$
will then be called the real vector space associated to $V^{A}.$ Setting
\begin{equation}
J(u)=ju,\,\,\,\,\,\,P^{+}(u)=e^{+}u,\,\,\,P^{-}(u)=e^{-}u\,\,,\,\,\,u\in
V^{A},  \label{2.6}
\end{equation}

the expressions
\begin{equation}
\begin{array}{l}
J^{2}=1_{V},\,\,\,\,P^{+2}=P^{+},\,\,\,\,P^{-2}=P^{-},\,\,\,\,\,\,P^{+}\circ
\,\,\,P^{-}=P^{-}\circ P^{+}=0 \\
P^{+}+P^{-}=1_{V},\,\,\,P^{+}-\,\,\,P^{-}=J, \\
P^{-}=(1/2)(1_{V}-J),\,\,\,\,\,\,P^{+}=(1/2)(1_{V}+J), \\
j^{2}=1,\,\,\,\,e^{+2}=e^{+},\,\,\,\,e^{-2}=e^{-},\,\,\,\,\,\,e^{+}\circ
\,\,\,e^{-}=e^{-}\circ e^{+}=0,\, \\
\,\,e^{+}+e^{-}=1,\,\,\,e^{+}-\,\,\,e^{-}=j, \\
e^{-}=(1/2)(1-j),\,e^{+}=(1/2)(1+j).%
\end{array}%
\end{equation}

can be written. Also, it is found that
\begin{equation}
\begin{array}{l}
J(\frac{\partial }{\partial x^{i}})=\frac{\partial }{\partial y^{i}}=\mathbf{%
j}\frac{\partial }{\partial x^{i}},\,\,J(\frac{\partial }{\partial y^{i}})=%
\frac{\partial }{\partial x^{i}}=\mathbf{j}\frac{\partial }{\partial y^{i}}%
\,, \\
P^{\mp }(\frac{\partial }{\partial z^{i}})=\mathbf{-}e^{\mp }\frac{\partial
}{\partial z^{i}},\,P^{\mp }(\frac{\partial }{\partial \overline{z}^{i}}%
)=e^{\mp }\frac{\partial }{\partial \overline{z}^{i}}, \\
P^{*\mp }(dz^{i})=\mathbf{-}e^{\mp }dz^{i},P^{*\mp }(d\overline{z}%
^{i})=e^{\mp }d\overline{z}^{i}.%
\end{array}
\label{2.8}
\end{equation}

\textbf{Definition 2: }An \textit{almost bi-para}-\textit{Hermitian manifold
}$(M,P^{+},\,P^{-},D_{1},D_{2})$ is a differentiable manifold $M$ with an
almost product structure $(P^{+},P^{-})$ and a pseudo-Riemannian metric $g$,
compatible in the sense that
\begin{equation}
g((P^{+}-P^{-})(X),Y)+g(X,(P^{+}-P^{-})(Y))=0,\,\,\,\,\forall X,Y\in \chi
(D_{1})(or\,\,\,\,\,\,\chi (D_{2})).  \label{2.9}
\end{equation}

A \textit{bi}-\textit{para}-\textit{Hermitian manifold }is a manifold with
an integrable almost bi-para-Hermitian structure $(g,P^{+},P^{-})$. Let us
take an almost bi-para-Hermitian manifold $(M,P^{+},\,P^{-},D_{1},D_{2}).$
Then, 2-covariant skew tensor field $\Phi $ defined by $\Phi
(X,Y)=g(X,(P^{+}-P^{-})(Y))$ is called \textit{fundamental 2-form}$.$ An
almost bi-para-Hermitian manifold $(M,P^{+},P^{-},D_{1},D_{2})$ is so-called
as an \textit{almost bi-para-K\"{a}hlerian manifold} $M$ if $\Phi $ is
closed. A bi-para-Hermitian manifold $(M,P^{+},P^{-},D_{1},D_{2})$ can be
said to be a \textit{bi}-\textit{para-K\"{a}hlerian manifold} $M$ if $\Phi $
is closed.

\section{Bi-Para-Lagrangians}

In this section, bi-para-Euler-Lagrange equations and a bi-para-mechanical
system can be obtained for classical mechanics structured under the
consideration of the basis $\{e^{+},e^{-}\}$ on Lagrangian distributions $%
D_{1}$ and $D_{2}$ of bi-para-K\"{a}hlerian manifold $M.$

Let $(P^{+},P^{-})$ be an almost bi-para-complex structure on the bi-para-K%
\"{a}hlerian manifold $M,$ and $(z^{i},\overline{z}^{i})$ be its complex
structures. Let semispray be the vector field $\xi $ given by

\begin{equation}
\begin{array}{l}
\xi =e^{+}(\xi ^{i+}\frac{\partial }{\partial z^{i}}+\overline{\xi }^{i+}%
\frac{\partial }{\partial \overline{z}^{i}})+e^{-}(\xi ^{i-}\frac{\partial }{%
\partial z^{i}}+\overline{\xi }^{i-}\frac{\partial }{\partial \overline{z}%
^{i}}); \\
z^{i}=\,z^{i+}e^{+}+z^{i-}e^{-};\,\overset{.}{\,z}^{i}=\overset{.}{\,z}%
^{i+}e^{+}+\overset{.}{z}^{i-}e^{-}=\xi ^{i+}e^{+}+\xi ^{i-}e^{-}; \\
\overline{z}^{i}=\,\overline{z}^{i+}e^{+}+\overline{z}^{i-}e^{-};\,\overset{.%
}{\,\overline{z}}^{i}=\overset{.}{\,\overline{z}}^{i+}e^{+}+\overset{.}{%
\overline{z}}^{i-}e^{-}=\overline{\xi }^{i+}e^{+}+\overline{\xi }^{i-}e^{-};%
\end{array}
\label{3.1}
\end{equation}
where the dot indicates the derivative with respect to time $t$. The vector
field denoted by $V=(P^{+}-P^{-})(\xi )$ and given by
\begin{equation}
(P^{+}-P^{-})(\xi )=e^{+}(-\xi ^{i+}\frac{\partial }{\partial z^{i}}+%
\overline{\xi }^{i+}\frac{\partial }{\partial \overline{z}^{i}})+e^{-}(\xi
^{i-}\frac{\partial }{\partial z^{i}}-\overline{\xi }^{i-}\frac{\partial }{%
\partial \overline{z}^{i}})  \label{3.2}
\end{equation}
is called \textit{bi}-\textit{para}-\textit{Liouville vector field} on the
bi-para-K\"{a}hlerian manifold $M$. The maps given by $T,P:M\rightarrow A$
such that $T=\frac{1}{2}m_{i}(\overline{z}^{i})^{2}=\frac{1}{2}m_{i}(\overset%
{.}{z}^{i})^{2},P=m_{i}gh$ are called \textit{the kinetic energy} and
\textit{the potential energy of the system,} respectively.\textit{\ }Here%
\textit{\ }$m_{i},g$ and $h$ stand for mass of a mechanical system having $m$
particles, the gravity acceleration and distance to the origin of a
mechanical system on the bi-para-K\"{a}hlerian manifold, respectively. Then $%
L:M\rightarrow A$ is a map that satisfies the conditions; i) $L=T-P$ is a
\textit{bi-para}-\textit{Lagrangian function, ii)} the function given by $%
E_{L}=V(L)-L$ is \textit{a bi-para-energy function}.

The operator $i_{(P^{+}-P^{-})}$ induced by $P^{+}-P^{-}$ and shown by
\begin{equation}
i_{P^{+}-P^{-}}\omega (Z_{1},Z_{2},...,Z_{r})=\sum_{i=1}^{r}\omega
(Z_{1},...,(P^{+}-P^{-})Z_{i},...,Z_{r})  \label{3.3}
\end{equation}
is said to be \textit{vertical derivation, }where $\omega \in \wedge
^{r}{}M, $ $Z_{i}\in \chi (M).$ The \textit{vertical differentiation} $%
d_{(P^{+}-P^{-})}$ is defined by
\begin{equation}
d_{(P^{+}-P^{-})}=[i_{(P^{+}-P^{-})},d]=i_{(P^{+}-P^{-})}d-di_{(P^{+}-P^{-})}
\label{3.4}
\end{equation}
where $d$ is the usual exterior derivation. For an almost para-complex
structure $P^{+}-P^{-}$, the closed bi-para-K\"{a}hlerian form is the closed
2-form given by $\Phi _{L}=-dd_{_{(P^{+}-P^{-})}}L$ such that
\begin{equation}
d_{_{(P^{+}-P^{-})}}=e^{+}B-e^{-}B:\mathcal{F}(M)\rightarrow \wedge ^{1}{}M
\label{3.5}
\end{equation}
where
\[
B=-\frac{\partial }{\partial z^{i}}dz^{i}+\frac{\partial }{\partial
\overline{z}^{i}}d\overline{z}^{i}.
\]
Then
\begin{equation}
\Phi _{L}=e^{+}C-e^{-}C  \label{3.6}
\end{equation}

where
\[
C=\frac{\partial ^{2}L}{\partial z^{j}\partial z^{i}}dz^{j}\wedge dz^{i}%
\mathbf{-}\frac{\partial ^{2}L}{\partial z^{j}\partial \overline{z}^{i}}%
dz^{j}\wedge d\overline{z}^{i}+\frac{\partial ^{2}L}{\partial \overline{z}%
^{j}\partial z^{i}}d\overline{z}^{j}\wedge dz^{i}-\frac{\partial ^{2}L}{%
\partial \overline{z}^{j}\partial \overline{z}^{i}}d\overline{z}^{j}\wedge d%
\overline{z}^{i}.
\]
Let $\xi $ be the second order differential equations given by \textbf{Eq. }(%
\ref{1.1}) and
\begin{equation}
\begin{array}{l}
i_{\xi }\Phi _{L}=\Phi _{L}(_{\xi })=e^{+}[\xi ^{i+}\frac{\partial ^{2}L}{%
\partial z^{j}\partial z^{i}}\delta _{i}^{j}dz^{i}-\xi ^{i+}\frac{\partial
^{2}L}{\partial z^{j}\partial z^{i}}dz^{j}-\xi ^{i+}\frac{\partial ^{2}L}{%
\partial z^{j}\partial \overline{z}^{i}}\delta _{i}^{j}d\overline{z}^{i}+%
\overline{\xi }^{i+}\frac{\partial ^{2}L}{\partial z^{j}\partial \overline{z}%
^{i}}dz^{j} \\
\mathbf{-}\xi ^{i+}\frac{\partial ^{2}L}{\partial \overline{z}^{j}\partial
z^{i}}d\overline{z}^{j}+\overline{\xi }^{i+}\frac{\partial ^{2}L}{\partial
\overline{z}^{j}\partial z^{i}}\delta _{i}^{j}dz^{i}-\overline{\xi }^{i+}%
\frac{\partial ^{2}L}{\partial \overline{z}^{j}\partial \overline{z}^{i}}%
\delta _{i}^{j}d\overline{z}^{i}+\overline{\xi }^{i+}\frac{\partial ^{2}L}{%
\partial \overline{z}^{j}\partial \overline{z}^{i}}d\overline{z}^{j}] \\
e^{-}[-\xi ^{i-}\frac{\partial ^{2}L}{\partial z^{j}\partial z^{i}}\delta
_{i}^{j}dz^{i}+\xi ^{i-}\frac{\partial ^{2}L}{\partial z^{j}\partial z^{i}}%
dz^{j}+\xi ^{i-}\frac{\partial ^{2}L}{\partial z^{j}\partial \overline{z}^{i}%
}\delta _{i}^{j}d\overline{z}^{i}-\overline{\xi }^{i-}\frac{\partial ^{2}L}{%
\partial z^{j}\partial \overline{z}^{i}}dz^{j} \\
\mathbf{+}\xi ^{i-}\frac{\partial ^{2}L}{\partial \overline{z}^{j}\partial
z^{i}}d\overline{z}^{j}-\overline{\xi }^{i-}\frac{\partial ^{2}L}{\partial
\overline{z}^{j}\partial z^{i}}\delta _{i}^{j}dz^{i}+\overline{\xi }^{i-}%
\frac{\partial ^{2}L}{\partial \overline{z}^{j}\partial \overline{z}^{i}}%
\delta _{i}^{j}d\overline{z}^{i}-\overline{\xi }^{i-}\frac{\partial ^{2}L}{%
\partial \overline{z}^{j}\partial \overline{z}^{i}}d\overline{z}^{j}].%
\end{array}
\label{3.7}
\end{equation}
Since the closed bi-para-K\"{a}hlerian form $\Phi _{L}$ on $M$ is in the
structure of bi-para-symplectic, it is found

\begin{equation}
\begin{array}{l}
E_{L}=e^{+}(-\xi ^{i+}\frac{\partial L}{\partial z^{i}}+\overline{\xi }^{i+}%
\frac{\partial L}{\partial \overline{z}^{i}})+e^{-}(\xi ^{i-}\frac{\partial L%
}{\partial z^{i}}-\overline{\xi }^{i-}\frac{\partial L}{\partial \overline{z}%
^{i}})-L%
\end{array}
\label{3.8}
\end{equation}
and thus
\begin{equation}
\begin{array}{l}
dE_{L}=-\xi ^{i+}\frac{\partial ^{2}L}{\partial z^{j}\partial z^{i}}%
dz^{j}e^{+}+\overline{\xi }^{i+}\frac{\partial ^{2}L}{\partial z^{j}\partial
\overline{z}^{i}}dz^{j}e^{+}\mathbf{-}\xi ^{i+}\frac{\partial ^{2}L}{%
\partial \overline{z}^{j}\partial z^{i}}d\overline{z}^{j}e^{+}+\overline{\xi
}^{i+}\frac{\partial ^{2}L}{\partial \overline{z}^{j}\partial \overline{z}%
^{i}}d\overline{z}^{j}e^{+} \\
-\frac{\partial ^{2}L}{\partial z^{j}}dz^{i}+\xi ^{i-}\frac{\partial ^{2}L}{%
\partial z^{j}\partial z^{i}}dz^{j}e^{-}-\overline{\xi }^{i-}\frac{\partial
^{2}L}{\partial z^{j}\partial \overline{z}^{i}}dz^{j}e^{-}\mathbf{+}\xi ^{i-}%
\frac{\partial ^{2}L}{\partial \overline{z}^{j}\partial z^{i}}d\overline{z}%
^{j}e^{-} \\
-\overline{\xi }^{i-}\frac{\partial ^{2}L}{\partial \overline{z}^{j}\partial
\overline{z}^{i}}d\overline{z}^{j}e^{-}-\frac{\partial ^{2}L}{\partial
\overline{z}^{j}}d\overline{z}^{i}.%
\end{array}
\label{3.9}
\end{equation}
With the use of \textbf{Eq.} (\ref{1.1}), the following expression can be
obtained:

\begin{equation}
\begin{array}{l}
\xi ^{i+}\frac{\partial ^{2}L}{\partial z^{j}\partial z^{i}}dz^{j}e^{+}-\xi
^{i+}\frac{\partial ^{2}L}{\partial z^{j}\partial \overline{z}^{i}}d%
\overline{z}^{j}e^{+}+\overline{\xi }^{i+}\frac{\partial ^{2}L}{\partial
\overline{z}^{j}\partial z^{i}}dz^{j}e^{+}-\overline{\xi }^{i+}\frac{%
\partial ^{2}L}{\partial \overline{z}^{j}\partial \overline{z}^{i}}d%
\overline{z}^{j}e^{+} \\
-\xi ^{i-}\frac{\partial ^{2}L}{\partial z^{j}\partial z^{i}}dz^{j}e^{-}+\xi
^{i-}\frac{\partial ^{2}L}{\partial z^{j}\partial \overline{z}^{i}}d%
\overline{z}^{j}e^{-}-\overline{\xi }^{i-}\frac{\partial ^{2}L}{\partial
\overline{z}^{j}\partial z^{i}}dz^{j}e^{-}+\overline{\xi }^{i-}\frac{%
\partial ^{2}L}{\partial \overline{z}^{j}\partial \overline{z}^{i}}d%
\overline{z}^{j}e^{-}+\frac{\partial ^{2}L}{\partial z^{j}}dz^{i}+\frac{%
\partial ^{2}L}{\partial \overline{z}^{j}}d\overline{z}^{i}=0.%
\end{array}
\label{3.10}
\end{equation}
If a curve denoted by $\alpha :A\rightarrow M$ is considered to be an
integral curve of $\xi ,$ then the equations given in the following are
\begin{equation}
\begin{array}{l}
\left[ \xi ^{i+}\frac{\partial ^{2}L}{\partial z^{j}\partial z^{i}}+%
\overline{\xi }^{i+}\frac{\partial ^{2}L}{\partial \overline{z}^{j}\partial
z^{i}}\right] dz^{j}e^{+}-\left[ \xi ^{i-}\frac{\partial ^{2}L}{\partial
z^{j}\partial z^{i}}+\overline{\xi }^{i-}\frac{\partial ^{2}L}{\partial
\overline{z}^{j}\partial z^{i}}\right] dz^{j}e^{-} \\
-\left[ \xi ^{i+}\frac{\partial ^{2}L}{\partial z^{j}\partial \overline{z}%
^{i}}+\overline{\xi }^{i+}\frac{\partial ^{2}L}{\partial \overline{z}%
^{j}\partial \overline{z}^{i}}\right] d\overline{z}^{j}e^{+}+\left[ \xi ^{i-}%
\frac{\partial ^{2}L}{\partial z^{j}\partial \overline{z}^{i}}+\overline{\xi
}^{i-}\frac{\partial ^{2}L}{\partial \overline{z}^{j}\partial \overline{z}%
^{i}}\right] d\overline{z}^{j}e^{-}+\frac{\partial L}{\partial z^{j}}dz^{j}+%
\frac{\partial L}{\partial \overline{z}^{j}}d\overline{z}^{j}=0%
\end{array}
\label{3.11}
\end{equation}
or alternatively
\[
\begin{array}{l}
(e^{+}-e^{-})(\left[ \xi ^{i+}\frac{\partial }{\partial z^{j}{}^{i}}+%
\overline{\xi }^{i+}\frac{\partial }{\partial \overline{z}^{j}}\right] e^{+}+%
\left[ \xi ^{i-}\frac{\partial }{\partial z^{j}}+\overline{\xi }^{i-}\frac{%
\partial }{\partial \overline{z}^{j}}\right] e^{-})(\frac{\partial L}{%
\partial z^{i}})dz^{j} \\
(e^{+}-e^{-})(-\left[ \xi ^{i+}\frac{\partial }{\partial z^{j}{}^{i}}+%
\overline{\xi }^{i+}\frac{\partial }{\partial \overline{z}^{j}}\right] e^{+}-%
\left[ \xi ^{i-}\frac{\partial }{\partial z^{j}}+\overline{\xi }^{i-}\frac{%
\partial }{\partial \overline{z}^{j}}\right] e^{-})(\frac{\partial L}{%
\partial \overline{z}^{i}})d\overline{z}^{j}+\frac{\partial L}{\partial z^{j}%
}dz^{j}+\frac{\partial L}{\partial \overline{z}^{j}}d\overline{z}^{j}=0.%
\end{array}
\]

Then the following equations are found:
\begin{equation}
\begin{array}{l}
(e^{+}-e^{-})\frac{\partial }{\partial t}\left( \frac{\partial L}{\partial
z^{i}}\right) +\frac{\partial L}{\partial z^{i}}=0,\,\,\,\,-(e^{+}-e^{-})%
\frac{\partial }{\partial t}\left( \frac{\partial L}{\partial \overline{z}%
^{i}}\right) +\frac{\partial L}{\partial \overline{z}^{i}}=0%
\end{array}
\label{3.12}
\end{equation}

or
\begin{equation}
\begin{array}{l}
(e^{+}-e^{-})\frac{\partial }{\partial t}\left( \frac{\partial L}{\partial
z^{i}}\right) +\frac{\partial L}{\partial z^{i}}=0,\,\,\,\,\,\,\,%
\,(e^{+}-e^{-})\frac{\partial }{\partial t}\left( \frac{\partial L}{\partial
\overline{z}^{i}}\right) -\frac{\partial L}{\partial \overline{z}^{i}}=0.%
\end{array}
\label{3.13}
\end{equation}
Thus the equations obtained in \textbf{Eq. }(\ref{3.13}) are seen to be a
\textit{bi-para-Euler-Lagrange equations} on the distributions $D_{1}$ and $%
D_{2},$ and then the triple $(M,\Phi _{L},\xi )$ is seen to be a \textit{bi}-%
\textit{para-mechanical system }with taking into account the basis $%
\{e^{+},e^{-}\}$ on the distributions $D_{1}$ and $D_{2}$.

\section{Bi-Para-Hamiltonians}

Here, bi-para-Hamiltonian equations and bi-para-Hamiltonian mechanical
system for classical mechanics structured on the distributions ${}D_{1}$ and
$D_{2}$ are derived.

Let $M$ be a bi-para-K\"{a}hlerian manifold and $\left\{ z_{i},\overline{z}%
_{i}\right\} $ be its complex coordinates. Let $\{\frac{\partial }{\partial
z_{i}}|_{p},\frac{\partial }{\partial \overline{z}_{i}}|_{p}\}$ and $%
\{\left. dz_{i}\right| _{p},\left( d\overline{z}_{i}\right) _{p}\},$ also,
be bases over para-complex number $A$ of tangent space $T_{p}(M)$ and
cotangent space $T_{p}^{*}(M)$ of $M,$ respectively$.$ Let us assume that an
almost bi-para-complex structure, a bi-para-Liouville form and a
bi-para-complex 1-form on the distributions ${}D_{1}$ and $D_{2}$ are shown
by $P^{*+}-P^{*-}$, $\lambda $ and $\omega $, respectively$.$ Then $\omega =%
\frac{1}{2}[(z_{i}d\overline{z}_{i}+\overline{z}_{i}dz_{i})e^{+}+(z_{i}d%
\overline{z}_{i}+\overline{z}_{i}dz_{i})e^{-}]$ and $\lambda
=(P^{*+}-P^{*-})(\omega )=\frac{1}{2}[(z_{i}d\overline{z}_{i}-\overline{z}%
_{i}dz_{i})e^{+}-(z_{i}d\overline{z}_{i}-\overline{z}_{i}dz_{i})e^{-}].$ It
is concluded that if $\Phi $ is a closed bi-para-K\"{a}hlerian form on the
distributions ${}D_{1}$ and $D_{2},$ then $\Phi $ is also a
bi-para-symplectic structure on ${}D_{1}$ and $D_{2}$.

Consider that bi-para-Hamiltonian vector field $Z_{H}$ associated with
bi-para-Hamiltonian energy $H$ is given by
\begin{equation}
Z_{H}=(Z_{i}\frac{\partial }{\partial z_{i}}+\overline{Z}_{i}\frac{\partial
}{\partial \overline{z}_{i}})e^{+}+(Z_{i}\frac{\partial }{\partial z_{i}}+%
\overline{Z}_{i}\frac{\partial }{\partial \overline{z}_{i}})e^{-}.
\label{4.2}
\end{equation}

Then
\[
\Phi =-d\lambda =(e^{+}-e^{-})(d\overline{z}_{i}\wedge dz_{i})
\]
and
\begin{equation}
i_{Z_{H}}\Phi =\Phi (Z_{H})=(\overline{Z}_{i}dz_{i}\mathbf{-}Z_{i}d\overline{%
z}_{i})e^{+}+(-\overline{Z}_{i}dz_{i}\mathbf{+}Z_{i}d\overline{z}_{i})e^{-}.
\label{4.4}
\end{equation}
Moreover, the differential of bi-para-Hamiltonian energy is obtained as
follows:
\begin{equation}
dH=(\frac{\partial H}{\partial z_{i}}dz_{i}+\frac{\partial H}{\partial
\overline{z}_{i}}d\overline{z}_{i})e^{+}+(\frac{\partial H}{\partial z_{i}}%
dz_{i}+\frac{\partial H}{\partial \overline{z}_{i}}d\overline{z}_{i})e^{-}.
\label{4.5}
\end{equation}
By means of \textbf{Eq.}(\ref{1.1}), using \textbf{Eq. }(\ref{4.4}) and
\textbf{Eq. }(\ref{4.5}), the bi-para-Hamiltonian vector field is found to
be
\begin{equation}
Z_{H}=(-\frac{\partial H}{\partial \overline{z}_{i}}\frac{\partial }{%
\partial z_{i}}+\frac{\partial H}{\partial z_{i}}\frac{\partial }{\partial
\overline{z}_{i}})e^{+}+(\frac{\partial H}{\partial \overline{z}_{i}}\frac{%
\partial }{\partial z_{i}}-\frac{\partial H}{\partial z_{i}}\frac{\partial }{%
\partial \overline{z}_{i}})e^{-}.  \label{4.6}
\end{equation}

Suppose that a curve
\begin{equation}
\alpha :I\subset A\rightarrow M  \label{4.7}
\end{equation}
be an integral curve of the bi-para-Hamiltonian vector field $Z_{H}$, i.e.,
\begin{equation}
Z_{H}(\alpha (t))=\overset{.}{\alpha },\,\,t\in I.  \label{4.8}
\end{equation}
In the local coordinates, it is obtained that
\begin{equation}
\alpha (t)=(z_{i}(t),\overline{z}_{i}(t))  \label{4.9}
\end{equation}
and
\begin{equation}
\overset{.}{\alpha }(t)=(\frac{dz_{i}}{dt}\frac{\partial }{\partial z_{i}}+%
\frac{d\overline{z}_{i}}{dt}\frac{\partial }{\partial \overline{z}_{i}}%
)e^{+}+(\frac{dz_{i}}{dt}\frac{\partial }{\partial z_{i}}+\frac{d\overline{z}%
_{i}}{dt}\frac{\partial }{\partial \overline{z}_{i}})e^{-}.  \label{4.10}
\end{equation}
Consideration of \textbf{Eqs. }(\ref{4.8}), (\ref{4.6}),\textbf{\ }(\ref%
{4.10}), the following results can be obtained
\begin{equation}
\frac{dz_{i}}{dt}e^{+}=-\frac{\partial H}{\partial \overline{z}_{i}}e^{+},%
\frac{dz_{i}}{dt}e^{-}=\frac{\partial H}{\partial \overline{z}_{i}}e^{-},%
\frac{d\overline{z}_{i}}{dt}e^{+}=\frac{\partial H}{\partial z_{i}}e^{+},%
\frac{d\overline{z}_{i}}{dt}e^{-}=-\frac{\partial H}{\partial z_{i}}e^{-}
\label{4.11}
\end{equation}
or in a more simplified form:

\begin{equation}
\frac{dz_{i}}{dt}=-(e^{+}-e^{-})\frac{\partial H}{\partial \overline{z}_{i}}%
,\,\,\,\,\frac{d\overline{z}_{i}}{dt}=(e^{+}-e^{-})\frac{\partial H}{%
\partial z_{i}}.  \label{4.12}
\end{equation}
Hence, the equations obtained in \textbf{Eq. }(\ref{4.12}) are seen to be
\textit{bi}-\textit{para Hamiltonian equations} on the distributions $%
{}D_{1} $ and $D_{2},$ and then the triple $(M,\Phi _{L},Z_{H})$ is seen to
be a \textit{bi-para-Hamiltonian mechanical system }with the use of basis $%
\{e^{+},e^{-}\}$ on the distributions ${}D_{1}$ and $D_{2}$.

\section{Conclusion}

This study has shown to exist physical proof of the mathematical equality
given by $M=D_{1}\oplus D_{2}.$ Also, formalisms of Lagrangian and
Hamiltonian mechanics have intrinsically been described with taking into
account the basis $\{e^{+},e^{-}\}$\ on Lagrangian distributions ${}D_{1}$\
and $D_{2}$\ of bi-para-K\"{a}hlerian manifold $M$.

Bi-para-Lagrangian and bi-para-Hamiltonian models arise to be a very
important tool since they present a simple method to describe the model for
bi-para-mechanical systems. In solving problems in classical mechanics, the
bi-para-complex mechanical system will then be easily usable model. With the
use of the corresponding approach, thus, a differential equation resulted in
mechanics is seen to have a non-trivial solution.

Since physical phenomena, as well-known, do not take place all over the
space, a new model for dynamical systems on subspaces is needed. Therefore,
equations (\ref{3.13}) and (\ref{4.12}) are only considered to be a first
step to realize how bi-para-complex geometry has been used in solving
problems in different physical area.

For further research, bi-para-complex Lagrangian and Hamiltonian vector
fields derived here are suggested to deal with problems in electrical,
magnetical and gravitational fields of physics.

\end{document}